\documentclass[a4paper,fleqn]{cas-dc}
\usepackage[numbers]{natbib}
\usepackage{subcaption}
\usepackage{booktabs}       
\usepackage{enumitem}
\newlist{noitemize}{itemize}{1}
\setlist[noitemize]{label={}, labelsep=0pt, leftmargin=0pt}
\usepackage{xspace}
\let\originalTextcolor\textcolor 
\renewcommand{\textcolor}[2]{\originalTextcolor{black}{#2}}
\makeatletter
\DeclareRobustCommand\onedot{\futurelet\@let@token\@onedot}
\def\@onedot{\ifx\@let@token.\else.\null\fi\xspace}

\def\eg{\emph{e.g}\onedot}

\def\etal{et al\onedot}
\makeatother
\usepackage{makecell}
\usepackage{arydshln}
\usepackage{booktabs}
\usepackage{multirow}
\begin{document}
\let\WriteBookmarks\relax
\def\floatpagepagefraction{1}
\def\textpagefraction{.001}

\shorttitle{\textcolor{red}{2.75D:} Boosting Learning by Representing 3D Imaging to 2D}

\shortauthors{Xin Wang, Ruisheng Su et~al.}

\title [mode = title]{\textcolor{red}{2.75D:} Boosting Learning by Representing 3D Medical Imaging to 2D Features for Small Data}                      



%
\author[1]{Xin Wang}[style=chinese]
\fnmark[1]
\affiliation[1]{
    organization={Department of Radiology, the Netherlands Cancer Institute},
    addressline={Plesmanlaan 121}, 
    city={Amsterdam},
    postcode={1066 CX}, 
    country={The Netherlands}
}

\author[2]{Ruisheng Su}[style=chinese]
\fnmark[1]
\affiliation[2]{
    organization={Erasmus Medical Center Rotterdam},
    addressline={Doctor Molewaterplein 40}, 
    city={Rotterdam},
    postcode={3015 CD}, 
    country={The Netherlands}
}

\author[3]{Weiyi Xie}[style=chinese]
\affiliation[3]{
    organization={Radboud University Medical Center},
    addressline={Geert Grooteplein Zuid 10},
    city={Nijmegen},
    postcode={6525 GA}, 
    country={The Netherlands}
}

\author[4]{Wenjin Wang}[style=chinese]
\affiliation[4]{
    organization={Biomedical Engineering Department of Southern University of Science and Technology},
    addressline={Xueyuan Blvd 1088},
    city={Shenzhen},
    postcode={518055}, 
    country={China}
}

\author[1]{Ritse Mann}

\author[5]{Jungong Han}[style=chinese]
\affiliation[5]{
    organization={Computer Science, Fellow of IAPR, Aberystwyth University},
    country={UK}
}

\author[1]{Tao Tan}[style=chinese]
\ead{taotanjs@gmail.com}
\cormark[1]

\cortext[cor1]{Corresponding author}
\fntext[fn1]{X. Wang and R. Su have equal contributions}



\begin{abstract}
In medical-data driven learning, 3D convolutional neural networks (CNNs) have started to show superior performance to 2D CNNs in numerous deep learning tasks, proving the added value of 3D spatial information in feature representation. However, the difficulty in collecting more training samples to converge, more computational resources and longer execution time make this approach less applied. Also, applying transfer learning on 3D CNN is challenging due to a lack of publicly available pre-trained 3D models. To tackle these issues, we proposed a novel 2D strategical representation of volumetric data, namely 2.75D. In this work, the spatial information of 3D images is captured in a single 2D view by a spiral-spinning technique. As a result, 2D CNN networks can also be used to learn volumetric information. Besides, we can fully leverage pre-trained 2D CNNs for downstream vision problems. We also explore a multi-view 2.75D strategy, 2.75D 3 channels (2.75D\textcolor{red}{$\times$}3), to boost the advantage of 2.75D. We evaluated the proposed methods on three public datasets with different modalities or organs (Lung CT, Breast MRI, and Prostate MRI), against their 2D, 2.5D, and 3D counterparts in classification tasks. 
\textcolor{red}{
Results show that the proposed methods significantly outperform other counterparts when all methods were trained from scratch on the lung dataset.
Such performance gain is more pronounced with transfer learning or in the case of limited training data.
Our methods also achieved comparable performance on other datasets.}
In addition, our \textcolor{red}{methods achieved} a substantial reduction in time consumption of training and inference \textcolor{red}{compared} with the 2.5D or 3D method.

\end{abstract}




\maketitle

\section{Introduction}
\subsection{Clinical Background}
Cancer is the second leading cause of death worldwide \cite{ebell2018cancer}, including lung, breast, and prostate cancers, which are the most widespread cancers in the developed world.
Early detection based on radiographic screening programs, such as X-ray, Computed Tomography (CT), and Magnetic Resonance Imaging (MRI), has the potential to improve the prognosis and reduce mortality for patients \citep{ebell2018cancer}. 
The effectiveness of cancer treatments heavily depends on the phase at which the cancer is diagnosed. 
For example, the five-year survival rate for a localized stage of non-small cell lung cancer is much higher than that for the distant stage \citep{blandin2017progress}. 
However, the implementation of widespread cancer screening programs in the world will likely lead to a substantial amount of medical images, while the workload and the shortage of screening radiologists are severe. 
In addition, due to the large size of medical images, finding early subtle abnormalities is a time-consuming and thus error-prone task for radiologists. 

\subsection{Related Work}
Benefited from the recent advances in Convolutional Neural Networks (CNNs) in various computer vision challenges \citep{russakovsky2015imagenet, simonyan2014very, he2016deep}, many algorithms for automatic cancer diagnosis on medical images have been proposed to help to improve radiologists' performance and reduce their workload. 
Logically, \textcolor{red}{increasingly} amount of studies have been devoted on the adoption of CNN in the medical imaging field, where data have different high-dimension natures, such as 2D images (X-ray \citep{tan2022multi}, 2D ultrasound \citep{xing2020lesion}) and 3D volumes (CT \citep{setio2016pulmonary}, \textcolor{red}{Computed Tomographic Angiography (CTA)} \citep{su2020automatic}, 3D ultrasound \citep{yang2019transferring}, MRI \citep{nielsen2018prediction}), \textcolor{red}{2D images+Time sequence (\citep{su2022spatio, zhang2022automated})}, 3D images+Time sequence (\citep{robben2020prediction}).

2D CNN is one of the most important networks in the machine learning field, which made impressive achievements in the past few years. 
Since AlexNet \citep{imagenet_cvpr09} was proposed, the revolution of deep CNN on large-scale datasets (\eg ImageNet \citep{imagenet_cvpr09}, which contains 1.4 million images with 1000 classes) was triggered. With emphasizing the importance of depth \citep{simonyan2014very}, deeper network structures emerged, such as GoogleNet \citep{szegedy2015going} and VGG \citep{simonyan2014very}, which significantly improved the performance in classification tasks. 
Afterward, He \etal~\cite{he2016deep} introduced skip connections in ResNet to solve the vanishing gradients problems.
While 2D CNNs go deeper and deeper, these state-of-the-art (SOTA) CNNs have also been extensively explored on medical images and different tasks, such as the malignancy classification of liver masses \citep{yasaka2018deep}, cancer types discrimination \citep{midya2018deep}, and intracranial aneurysm detection \citep{duan2019automatic}. 
Moreover, many well-known 2D CNNs pre-trained on the large ImageNet dataset are open to the public and readily accessible, which supports transfer learning. 
It is critical in the context of medical images with small datasets and a lack of annotations, due to the cost and necessary workload of radiologists. 


Regular 2D CNN approaches are well known for handling 2D image classification tasks. However, 2D CNNs are not directly applicable to 3D medical images. Based on the model input representation of 3D data, existing CNN methods can be mainly divided into three categories: 2D, 2.5D, and 3D. Firstly, a straightforward approach is to take a representative 2D slice from the 3D volume as model input \citep{jacobs2016computer}, named the 2D method. 
Such a representation generally fails to capture sufficient volumetric stereo information. Various studies have shown the inferior performance of 2D CNN versus other approaches which incorporate stereo information. For example, Setio \etal~\cite{setio2016pulmonary} utilized multiview 2D CNN for pulmonary nodule classification, which outperformed single-view 2D CNN. 

To best learn high dimensional features, many SOTA 3D CNN approaches have been investigated for 3D medical imaging \citep{dou2016multilevel, huang2017lung, ding2017accurate}, which can extract volumetric features from entire volumes without losing information, demonstrating promising improvements in many computer-aided diagnosis applications. 
For example, Yang \etal~\cite{yang2018visual} trained a 3D CNN for Alzheimer’s Disease classification. The superiority of 3D CNN was demonstrated in their work by comparing to baseline 2D CNNs. More examples of 3D CNN approaches in medical imaging include pulmonary nodule classification \citep{kang20173d}, brain hemorrhage classification \citep{jnawali2018deep}, Alzheimer diagnosis \citep{yang2018visual}, and breast cancer classification \citep{zhou2019weakly}.
However, due to the limited availability of data and high computational resources, practice of 3D CNNs on medical images is discouraged \citep{singh20203d}.


With the success of 2D CNNs and the limitations of 3D CNNs, various studies have also attempted to represent 3D volumes in multiple 2D views, which is the 2.5D method. For example, Lyksborg \etal~\cite{lyksborg2015ensemble} used 2D CNNs on three orthogonal 2D patches to incorporate 3D contextual information. Su \etal~\cite{su2015multi} utilized trained multiple CNN streams, each using one 2D view of the object for 3D shape classification. Similarly, Setio \etal~\cite{setio2016pulmonary} extracted nine differently oriented volume slices and fed them into separate streams of 2D CNNs for pulmonary nodule classification. 2.5D methods generally are between 2D and 3D approaches and better in training efficiency comparing to 3D methods \citep{setio2016pulmonary, nibali2017pulmonary,liu2017multiview, liu2018multi, xie2019automated}. As aforementioned, 2.5D CNN requires fewer GPU resources than 3D CNN, and incorporates richer volumetric information of neighboring voxels than the 2D method. On the other hand, the biggest problem from 2.5D is that the voxels from 3D are not strategically sampled in a single 2D image while the geometry relation between voxels is lost, which means 2.5D cannot strategically represent the information needed for diagnosis from 3D. 

Graph representation may also be able to represent the relations of different regions efficiently. For example, Zhang \etal~\cite{zhang2022automated} use a multiplex visibility graph to represent 2D+times series recordings. However, this method also has limitations, such as the need to correspond to a standard (brain) atlas to extract the graph structure. Su \etal~\cite{su2020autotici} defined the anatomical region by atlas-based registration and then extracted high-level features from CTA.
This is hard for lung, breast, or prostate imaging due to the huge personalized variations.


Is there a better 2D representation of 3D volumes other than 2.5D? To answer this, we revisit the spiral scanning technique in this work. Spiral scanning is an existing concept, which has been initially proposed in signal sampling in the physical space of imaging. For example, Wang \etal~\cite{wang2007approximate} adopted spiral scanning for cone-beam reconstruction in 2007. In the same year, spiral scanning was exploited for the sake of dynamic programming segmentation in 2D \citep{wang2007segmentation}. In 2014, Rana \etal~\cite{rana2014spiral} utilized spiral scanning to improve the control of atomic force microscope. Despite various applications of spiral scanning, it has not been explored for efficient 3D to 2D representation in CNNs. Recently, lots of efforts have been paid on aspects such as designing architectures of deep-learning networks, improving learning scheme by optimizing hyperparameters, and increasing data diversity by augmenting limited data. Rarely, research has been focused on the data representation domain. Data representation is more than image preprocessing. Preprocessing mainly refers to image normalization, noise reduction and artifact removal, which makes follow-up processing focus on clinical-related image features. Those transformations are pixel-level and intensity-level changes. Data representation further transforms the data into a new form which might make the follow-up processing be more comfortable.

In summary, the significant challenge in 3D medical imaging is the requirement for high-efficiency representation of volumetric information, beyond resource limitation and fully leveraging pre-trained CNNs. This is a key factor in the development of 3D medical imaging-based CAD systems in clinic practice. Current SOTA 3D approaches generally outperform 2D or 2.5D approaches while requiring significantly more data for training, consuming more GPU resources as well as limiting the ability to use transfer learning. 

\subsection{Contributions}
To this end, we proposed our 2.75D method\footnote{The TensorFlow and PyTorch implementations of our code available at \url{https://github.com/RuishengSu/2.75D}}, which boosts the computational efficiency and learning capability of deep learning networks on 3D image classification tasks using the spiral scanning technique. We have extended our previous Open Arxiv version \citep{su20202}, with explored robust multi-view 2.75D strategy and validated on more public datasets. Our approach serves as a generic framework for generating a sufficient 2D presentation of a 3D volume. When sample frequency is set to a small value, the approach can be reduced to 2D or 2.5 approaches. Meanwhile, our approach runs faster than 2.5D approaches, similar to the 2D method, while can maintain relatively good performance \textcolor{red}{compared} to 3D methods. By converting 3D volume to a sufficient 2D representation, it is beneficial to perform transfer learning using models trained over a large number of natural images \citep{imagenet_cvpr09}, whereas pre-trained models for 3D medical images are rare. The significant performance gain of incorporating transfer learning on 2.75D is demonstrated in comparison with 3D approaches.
The main contributions of this paper are as follows: 
\begin{itemize}[leftmargin=*]
\item We proposed a 2.75D strategy to extract efficient 2D representations of a 3D volume as the input to a 2D classification CNN instead of a 3D CNN.
\item We evaluated the proposed methods on three public datasets with different modalities and different organs and demonstrated its remarkable performance in comparison to 2D, 2.5D, and 3D approaches.
\item We explored the capability and advantage of 2.75D using transfer learning via pre-trained models from large-scale 2D image data sets that are publicly available.
\item We systematically investigated the effect of data size on training yield from different approaches.
\end{itemize}

\section{Method}\label{sec:method}
In this section, the proposed 2.75D strategies are explained in detail, especially how 3D volumes are transformed into corresponding 2D representations. Provided a 3D volume classification task, CNNs can be utilized to learn textural and volumetric information from the surrounding volume of a given candidate voxel. To boost the efficiency of such CNN based methods, we propose to first transform these 3D volumes into 2D representations using the spiral scanning technique, followed by a 2D CNN classifier which can learn volumetric information as in 3D CNNs. 

\begin{figure}[h]
\centering
\begin{subfigure}{\linewidth}
  \includegraphics[width=\textwidth]{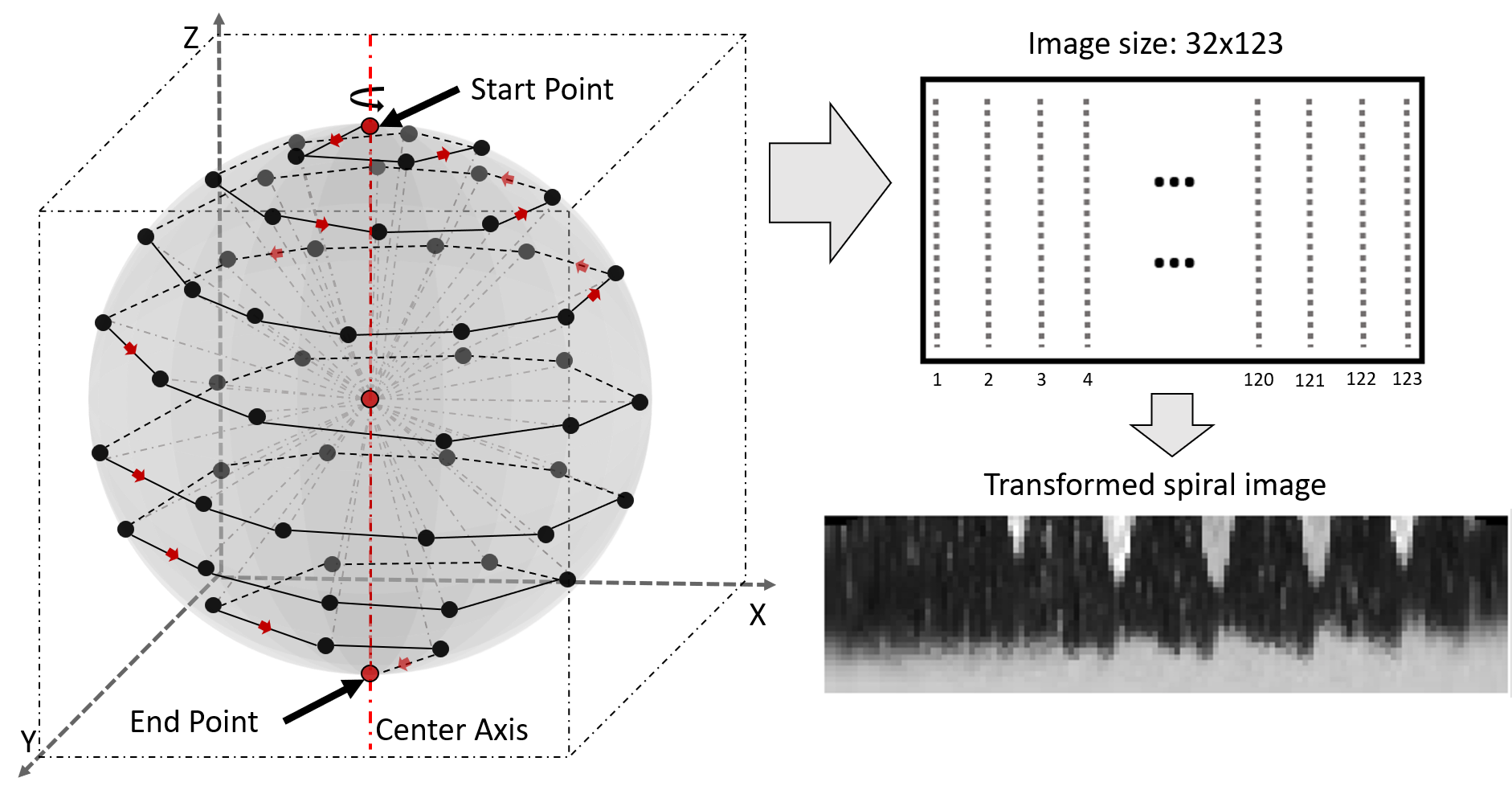}
  \caption{2.75D}
  \label{subfig:single_spiral_scan}
\end{subfigure}
\begin{subfigure}{\linewidth}
  \includegraphics[width=\textwidth]{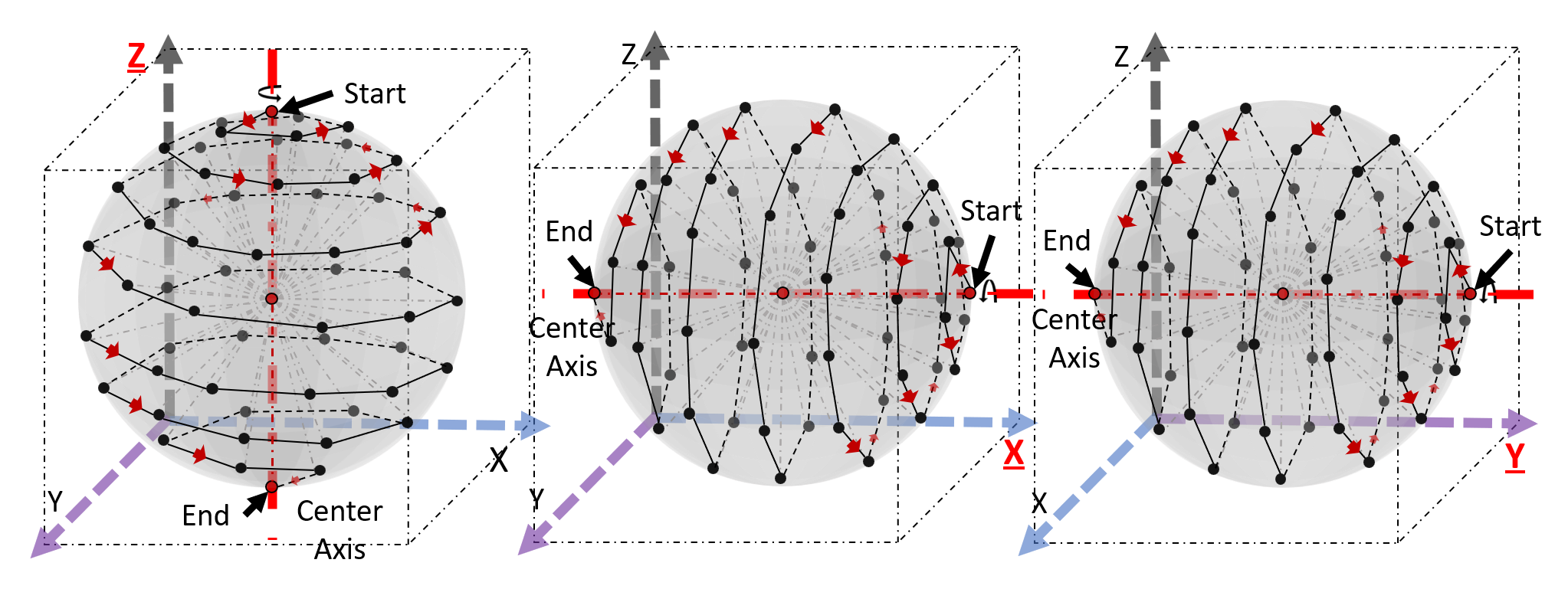}
  \caption{2.75D 3 channels (2.75D\textcolor{red}{$\times$}3)}
  \label{subfig:multi_spiral_scan}
\end{subfigure}
\caption{(a) Transformation of a 3D volume to 2D representation using spiral scanning. Each dotted line in light grey color represents a sampled radial line originated from the sphere center to a surface point. 32 intensity values are on each sampled radial line, which forms one column in the transformed 2D image. 123 radial lines from top to bottom of the sphere are ordered from left to right in the transformed 2D image. (b) Three different views for multi-view spiral scanning.}\label{fig:spiral_scan_flow}
\end{figure}


By representing 3D volumes in a 2D fashion, many existing 2D CNNs can be adopted for solving 3D image vision problems (e.g. nodule classification in chest CT, breast MRI, and prostate MRI). We hypothesize that such a 2D representation would contain volumetric information. For enhanced feature extraction from 3D volumes, we further explore a multi-spiral-view strategy named 2.75D 3channels (2.75D\textcolor{red}{$\times$}3). The patch extraction process is explained in detail below.

\begin{figure*}
  \centering
  \includegraphics[width=\textwidth]{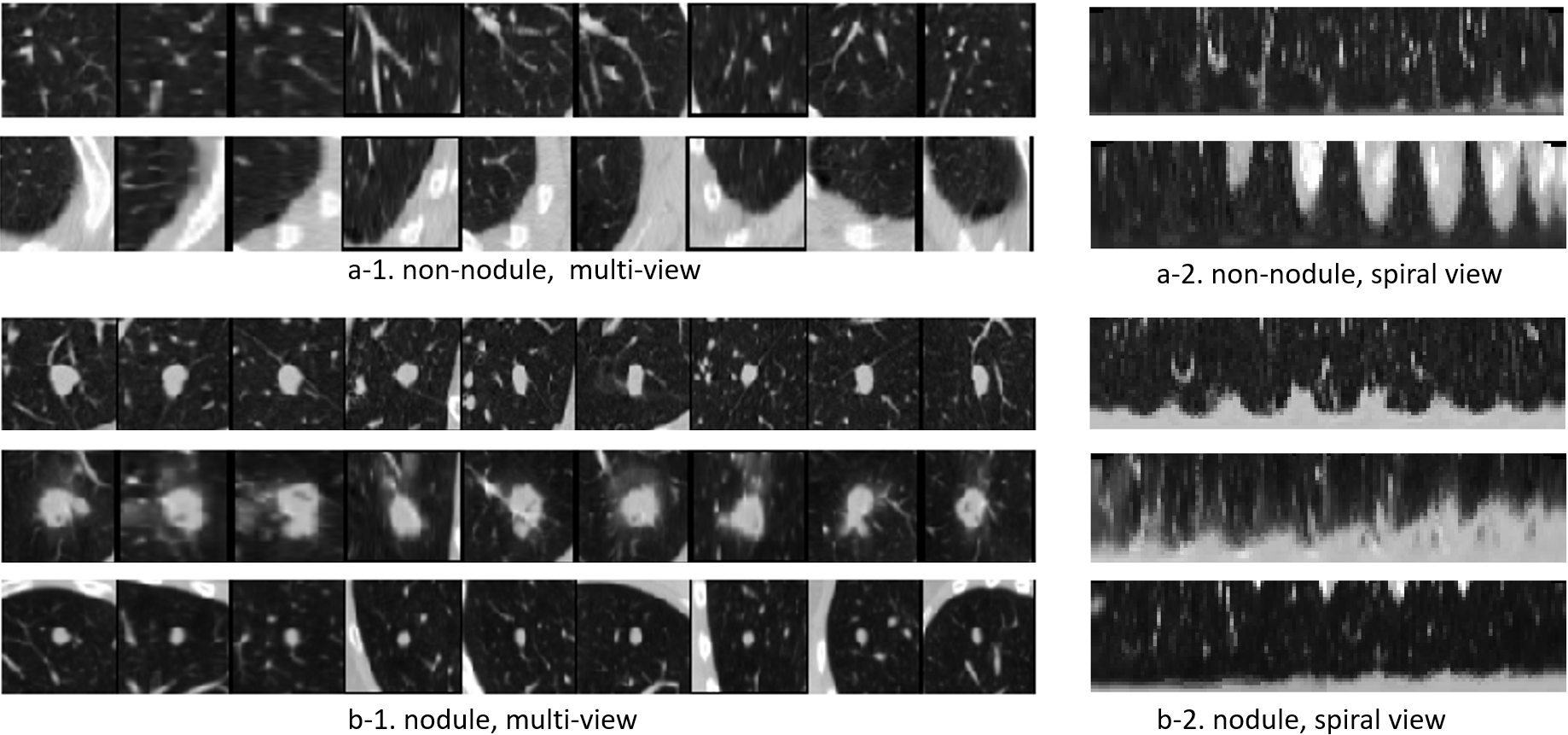}
  \caption{Patch examples from 2.5D and the proposed 2.75D in the LIDC-IDRI dataset. Left: nine differently oriented views of VOIs concatenated (size: 64\textcolor{red}{$\times$}(64\textcolor{red}{$\times$}9) = 64\textcolor{red}{$\times$}576 px); right: extracted spiral view patches of VOIs (size: 32\textcolor{red}{$\times$}123 px); a: non-nodule examples; b: nodule examples.}
  \label{fig:patch_examples}
\end{figure*}

\subsection{Spiral Scanning for 2D Patch Extraction}\label{spiral_scanning}
The workflow of 3D to 2D transformation is illustrated in Fig.~\ref{subfig:single_spiral_scan}. Given a 3D cubic volume of size $d\times d \times d$, a sphere with a radius of $r=d/2$ is defined at the volume center. The spiral scanning process sample voxels from 3D coordinate space to generate a 2D image via the movement of a scanning line (shown in gray dotted lines). During the scanning process, the starting point of the scanning line is always at the volume center, but its end-point moves on the sphere surface from the topmost point to the lowest point at the bottom, while circling around the central axis (default parallel to the Z-axis) counter-clockwise. \textcolor{red}{According to the patched 64 \textcolor{red}{$\times$} 64 \textcolor{red}{$\times$} 64 3D cubic, the heights of resampled images are always 32.} Such a movement on a 3D space can be decomposed into two rotation angles: the azimuth (longitude) angle $\alpha\in[0, \pi]$ defining the movement vertically and elevation (latitude) angle $\beta\in[0, 2\pi]$ describing the horizontal movement. To make the transform discrete, the spiral scanning undergoes a sampling process. First, the movement is captured in steps by sampling two rotation angles at a fixed interval. We show the sampled steps using a sequence of surface points (dark solid dots) along the scanning trajectory on the sphere surface (dark solid lines). Second, at each step, a fixed number of voxels are sampled along the scanning line in the 3D volume. All sampled voxels are rearranged in a 2D image sequentially as the sampling process goes, where rows indicate movement steps, and each column is filled with voxels sampled on the scanning line. 

Suppose that the spiral line is densely distributed around the sphere, the length of the spiral scanning orbit would be approximately equal to the sum of circumferences of a number of separate horizontal circles. For simplicity, the number of surface points is calculated based on this approximation. Suppose the angle step is $\pi/N$, the azimuth and elevation angles are evenly divided into $N$ ($\alpha_k=k\pi/N, k=0,1,2,...,N$) and $2N$ ($\beta_k=k\pi/N, k=0,1,2,...,2N$) sections, respectively. The number of sample points on a horizontal circle at $\alpha$ can be expressed as $2\pi|\sin(\alpha)|/(\pi/N)$, i.e., $2N\sin(\alpha)$, regardless of the sphere radius. Therefore, the total number of sample points on the sphere surface is approximately $4N^2/\pi$ when N is large, as expressed in Eq.\ref{eq:number_of_points}. \textcolor{red}{In this work, we choose $N=9$ for a volume of $64\times64\times64$ to produce 123 sample points that equal to the width of the resampled image (as shown in Fig.~\ref{subfig:single_spiral_scan}). More importantly, a $32\times123$ image is smaller than a 2D image ($64\times64$). When $N=10$, the number of sample points is 147, which will be larger than the 2D image. For a fair comparison with the 2D method and to prove that the superiority of our method is due to a higher efficient informative sampling strategy rather than more sample pixels, we set the image size to be $32\times123$. }


\begin{equation}\label{eq:number_of_points}
    \begin{split}
        & \sum_{k=0}^{N} 2N|\sin(k\pi/N)| \approx 2N \int_{0}^{N}|\sin(k\pi/N)|dk \\
        & = 2N\int_{0}^{N}(N/\pi)|\sin(k\pi/N)|d(k\pi/N) \\
        & = 2N^2/\pi \int_{0}^{\pi}|\sin x|dx \\
        & = 4N^2/\pi
    \end{split}
\end{equation}

As shown in Fig.~\ref{subfig:multi_spiral_scan}, the spiral scanning is around the center axis, which is parallel to the Z-axis by default. Similar to the 2.5D approach, when we use the x-axis and the y-axis, and the z-axis as the default axes, respectively, we can extract three 2D representations.

After spiral scanning, the 2D representations of the 3D volumes are obtained, which aggregate 3D information in all directions via the spiral scanning process. Some examples of the resulted spiral view images are shown in Fig.~\ref{fig:patch_examples}. These 2D patches can then be fed into neural networks for training and inference instead of 3D volumetric data.

\begin{figure*}[ht]
\centering
\begin{subfigure}{\linewidth}
  \centering\includegraphics[width=\textwidth]{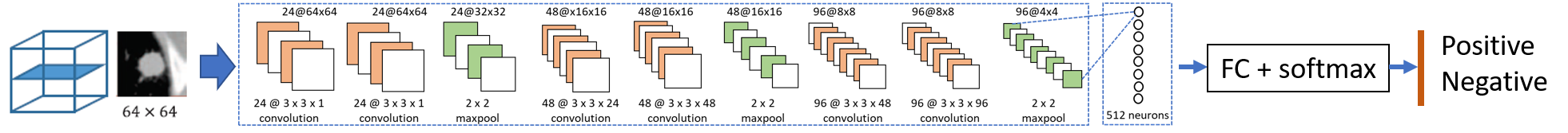}
  \caption{2D: single view 2D CNN architecture}
  \label{subfig:2D}
\end{subfigure}
\begin{subfigure}{\linewidth}
  \centering\includegraphics[width=\textwidth]{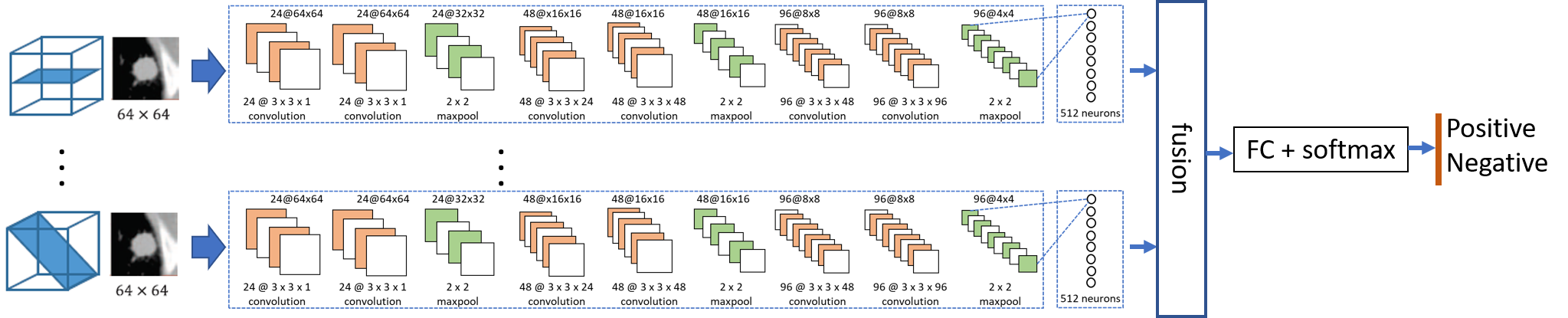}
  \caption{2.5D: multi-view 2D CNN architecture}
  \label{subfig:2.5D}
\end{subfigure}
\begin{subfigure}{\linewidth}
  \centering\includegraphics[width=\textwidth]{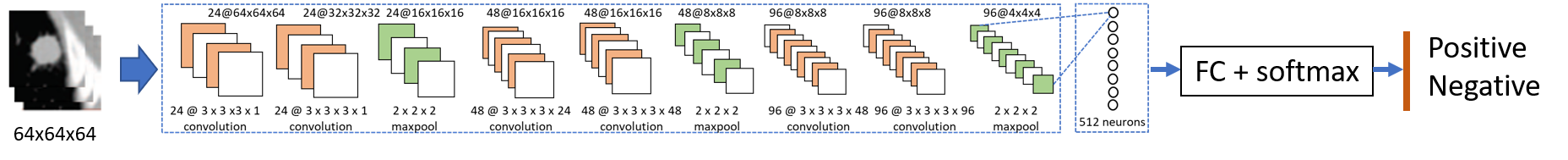}
  \caption{3D: volume based 3D CNN architecture}
  \label{subfig:3D}
\end{subfigure}
\begin{subfigure}{\linewidth}
  \centering\includegraphics[width=\textwidth]{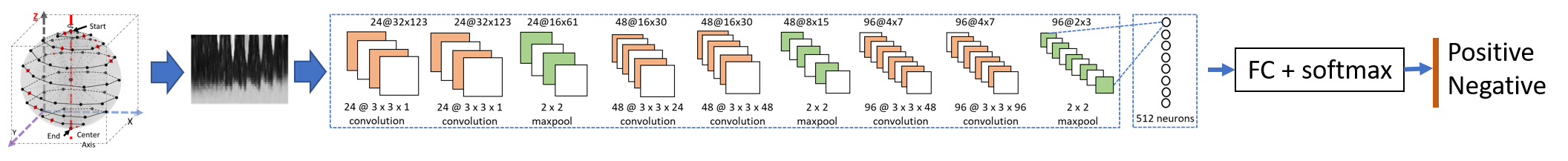}
  \caption{2.75D: spiral view based CNN architecture}
  \label{subfig:2.75D}
\end{subfigure}
\begin{subfigure}{\linewidth}
  \centering\includegraphics[width=\textwidth]{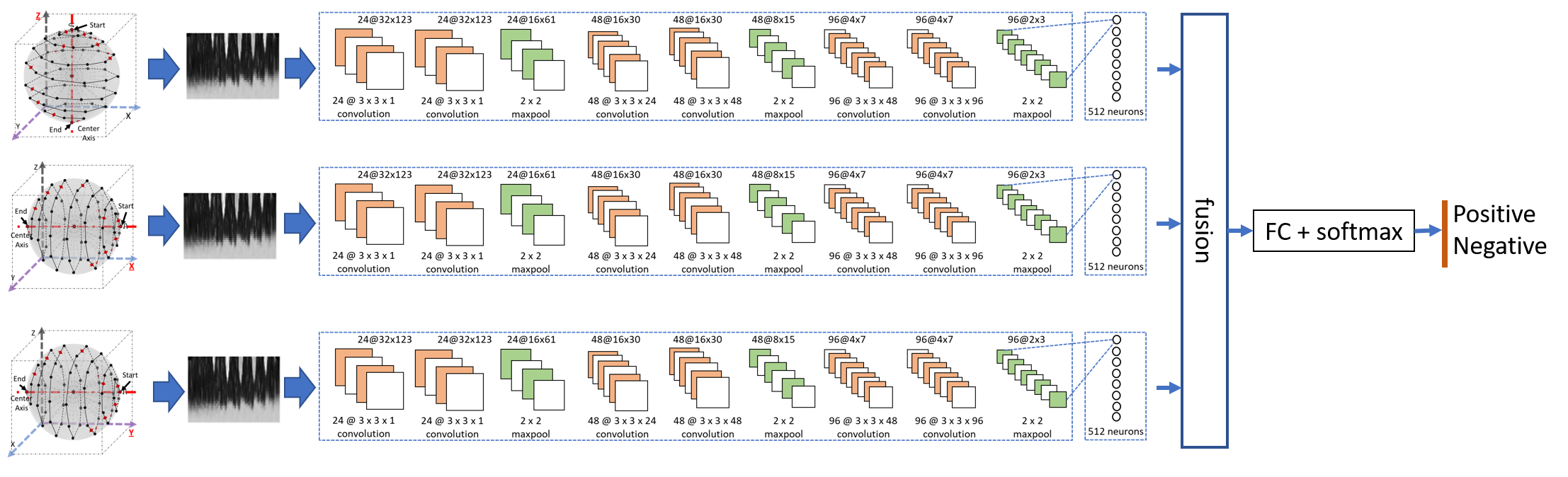}
  \caption{2.75D 3channels: multi-spiral-view based CNN architecture}
  \label{subfig:2.75Dx3}
\end{subfigure}
\caption{Architectures of 2D, 2.5D, 3D, 2.75D and 2.75D 3channels (2.75D\textcolor{red}{$\times$}3) approaches}\label{fig:architectures}
\end{figure*}

\section{Materials}\label{sec:materials}
To evaluate the performance of the proposed strategy, we compare the performance of four strategies (2D, 2.5D, 3D, 2.75D, and 2.75D 3 channels) on the classification task on three official data sets. They are from the LUNA16 nodule detection grand challenge \footnote{\url{https://luna16.grand-challenge.org/}}, the PROSTATEx challenge \footnote{\url{https://prostatex.grand-challenge.org/}} and Duke-Breast-Cancer-MRI \footnote{\url{https://doi.org/10.7937/TCIA.e3sv-re93}} \citep{saha2021dynamic}. We also demonstrate the effect of transfer learning on 2.75D. More, we evaluate the performance of different approaches with the adjustment of training data sizes. 
\subsection{Datasets and Preprocessing}

\subsubsection{LUNA16 - Nodule or non-nodule classification:}
The LUNA16 challenge consists of two sub-tasks based on a two-stage nodule detection pipeline. The first is to detect nodule candidates on given CT scans, which may include false positives; The second task is to classify each nodule candidate into nodule or non-nodule given its surrounding 3D volume on the CT scan. We participated in the false positive reduction in the nodule detection pipeline. In total, 888 CT scans are included in the challenge data set. CT scans all have slice thickness and are less than 2.5 \textcolor{red}{millimeters}. In the reference standard, nodules were annotated by the consensus of four radiologists in two phases. For the false positive reduction track, nodule candidates were generated using a set of existing automated nodule detection algorithms \citep{murphy2009large,jacobs2014automatic,setio2015automatic}. In total, 1120 out of 1186 nodules are detected with 551,065 candidates. More information regarding the annotation protocol and the procedure for nodule candidate generation can be found in \citep{armato2011lung}. 

\begin{table}[h]
\setlength\tabcolsep{0.5pt}
\scriptsize
  \centering
  \caption{Dataset distribution of positive and negative samples (including augmented)  in 10/5-fold cross-validation.}
  \begin{tabular}{cll:ll:ll}
    \toprule
    & \multicolumn{2}{l}{Lung CT(10-Fold)} &\multicolumn{2}{l}{Breast MRI(10-Fold)} & \multicolumn{2}{l}{Prostate MRI(5-Fold)} \\
    \cmidrule(r){2-3}  \cmidrule(rl){4-5} \cmidrule(rl){6-7}
    Subset  & \multicolumn{1}{l}{\makecell[l]{Nodule \\ (Augm.)}} & \multicolumn{1}{l}{\makecell[l]{Non- \\ nodule}} & \multicolumn{1}{l}{\makecell[l]{Cancer \\ (Augm.)}} &  \multicolumn{1}{l}{Normal} & \multicolumn{1}{l}{\makecell[l]{Malignant \\ (Augm.)}} &  \multicolumn{1}{l}{\makecell[l]{Benign \\ (Augm.)}}\\
    \midrule
    0   & 138 (78859) & 78997 & 49 (4900) & 4949 & 13 (5239) & 52 (5200) \\
    1   & 170 (70672) & 70842 & 49 (4900) & 4949 & 19 (5030) & 50 (4999) \\
    2   & 181 (74096) & 74277 & 49 (4900) & 4949 & 15 (5034) & 50 (4999) \\
    3   & 158 (75634) & 75792 & 48 (4800) & 4848 & 7 (5850) & 58 (5799) \\
    4   & 170 (76243) & 76413 & 49 (4900) & 4949 & 22 (4422) & 44 (4400) \\
    5   & 127 (75437) & 75564 & 49 (4899) & 4948 &   &    \\
    6   & 154 (76363) & 76517 & 48 (4800) & 4848 &   &    \\
    7   & 120 (74823) & 74943 & 48 (4800) & 4848 &   &    \\
    8   & 195 (74098) & 74293 & 48 (4800) & 4848 &   &    \\
    9   & 142 (72606) & 72748 & 48 (4800) & 4848 &   &    \\
    Total & 1555 (748831) & 750386 & 485 (48499) & 48984 & 76(25575) & 254(25397) \\
    \bottomrule
  \end{tabular}
  \label{table:dataset}
\end{table}

We followed the preprocessing procedure described by Setio \etal~\cite{setio2016pulmonary}. A 50\textcolor{red}{$\times$}50\textcolor{red}{$\times$}50 mm volume of interest (VOI) was extracted from each of the candidate nodule location. The size of VOI was chosen to ensure \textcolor{red}{full} visibility of all nodules ($<= 30 mm$) and sufficient context information. Each VOI was then resized into a 64\textcolor{red}{$\times$}64\textcolor{red}{$\times$}64 pixel 3D cube, resulting a resolution of 0.78 mm. Next, we rescaled the pixel intensity range from (-1000, 400 Houndsfield Unit (HU)) to (0,1) and clipped the intensity beyond this range. Then, we performed evaluation in 10-fold cross-validation across the selected 888 LIDC-IDRI \citep{armato2011lung} cases, which was split into \textcolor{red}{ten} subsets with the similar number of candidates in the Luna Challenge \citep{setio2017validation}. For each fold, we used seven subsets for training, two subsets for validation, and one subset for testing. The data size of each fold is shown in table \ref{table:dataset}. Specifically, the nodule sample size is 1555, whereas the non-nodule sample size is 750386, which is 482.5 times higher. To ensure a balanced dataset, we increased the number of nodules by randomly applying one or more of augmentation methods, which are explained in detail in Section \ref{subsubsec:augmentation}.

\subsubsection{Duke Breast Cancer MRI - Simulating cancer-normal classification:}
The Duke Breast Cancer MRI \textcolor{red}{dataset} is a single institutional, retrospective collection of 922 biopsy-confirmed invasive breast cancer patients \citep{saha2018machine} which contains a lot of clinical information (\eg{ demographic, clinical, pathology and treatment}). Radiologists also provided the annotations of the biggest lesion of every patient. We performed a cancer-normal classification task on this dataset. Specifically, the \textcolor{red}{tumours} in the MRI images were cropped, representing cancer samples. As for normal samples, 101 3D patches were randomly cropped from the bilateral breast regions away from the \textcolor{red}{tumour} region. Therefore bilateral breast cancer patients were excluded. For preprocessing, we resampled dynamic contrast-enhanced (DCE) MRI images to 1mm\textcolor{red}{$\times$}1mm\textcolor{red}{$\times$}1mm and rescaled the pixel intensity range to (0,1). Same as in the LUNA16, a 50\textcolor{red}{$\times$}50\textcolor{red}{$\times$}50 mm volume of interest (VOI) was extracted from each of the candidates, excluding the \textcolor{red}{tumours} larger than a patch. Each VOI was then resized into a 64\textcolor{red}{$\times$}64\textcolor{red}{$\times$}64 pixel 3D cube. Then 485 cases were randomly split into ten subsets with a similar number of candidates for the 10-fold cross-validation (Table \ref{table:dataset}). \textcolor{red}{For each fold, we used seven subsets for training, two subsets for validation, and one subset for testing.} We also employed random augmentation to balance the normal and cancer samples. 

\subsubsection{PROSTATEx - Lesion diagnostic classification:}
The details of PROSTATEx Challenge are described in \citep{armato2018prostatex}. The task is a diagnostic classification of clinically significant prostate lesions in prostate MRI images. In this study, a total of 330 lesions from DCE T1-weighted images of 204 patients were selected, as these images were provided with the ground truth of malignancy. We crop the lesions by providing the coordinate positions and also resize each VOI into a 64\textcolor{red}{$\times$}64\textcolor{red}{$\times$}64 pixel 3D cube. In order to ensure that the test set has enough samples for testing, we use five-fold cross-validation (Table \ref{table:dataset}). For each fold, we used three subsets for training, one subset for validation, and one subset for testing. The dataset was also oversampled and balanced by random augmentation in order to reduce the impact of the small and imbalanced dataset on the model.

\begin{figure*}[htb]
  \centering
  \includegraphics[clip, trim=0.4cm 0.3cm 0.4cm 0.3cm, width=.80\linewidth]{figures/data_aug.png}
\caption{Sapmles of different data augmentation.}
\label{fig:data_aug}
\end{figure*}

\subsection{Data Augmentation Pipeline}\label{subsubsec:augmentation}
Data augmentation is a widely used technique to increase the number of training and validation data samples to avoid imbalanced training and validation data sample sizes. A balanced and sufficiently large dataset is important for model robustness and overfitting prevention. In this experiment, we increased the number of samples by randomly applying the following augmentation steps before resampling: 1) rotation from $0^o$ to $360^o$ along one or two axes (randomly chosen) of the 64\textcolor{red}{$\times$}64\textcolor{red}{$\times$}64 pixel 3D volume; 2) flipping the volume along one random axis (X, Y, or Z); 3) zooming in along a random axis or axes to maximum 125\%; 4) translating the volume by maximum 25\% along a random axis, with zero padding. The volume size was kept the same during augmentation; 5) random Gaussian Noise; 6) random Gaussian Sharpen; 7) random Gaussian Smooth; 8) random Histogram Shift. The examples of results using different augmentation methods are shown in Fig.~\ref{fig:data_aug}. These random augmentations were implemented using MONAI (\url{https://monai.io/}) randomized data augmentation transforms. 

\section{Experiments}\label{sec:experiment}
After preprocessing and 3D data augmentation, three datasets with sample volumes of size 64\textcolor{red}{$\times$}64\textcolor{red}{$\times$}64 are obtained. These datasets are used to evaluate the proposed 2.75D and 2.75D\textcolor{red}{$\times$}3 against other existing strategies. Depending on the strategy, either the entire 3D volume or a portion of it is fed into CNNs for nodule classification. In this experiment, 2.75D and 2.75D\textcolor{red}{$\times$}3 were assessed against \textcolor{red}{their} 2D, 2.5D, and 3D counterparts. 

\subsection{Evaluation of CNN Strategies}\label{experiment:strategies}

\begin{table}[h]
\setlength\tabcolsep{3.5pt}
\scriptsize
  \centering
  \caption{Comparison of model size and computational cost expressed in terms of Parameters (Mb), \textcolor{red}{FLOPs (Gbps)}, and Inference Speed (images/s), respectively. Note that, in order to perform a fair comparison, the inference time includes resampling a 64\textcolor{red}{$\times$}64\textcolor{red}{$\times$}64 cube to a 2D, 2.5D, 2.75D, or 2.75D\textcolor{red}{$\times$}3 images. Input patches are resized to fit pre-trained models in case of transfer learning. }
  \begin{tabular}{ll:ll:ll:l}
    \toprule
    \multicolumn{1}{l}{Strategy} & \multicolumn{1}{l}{\makecell[l]{Patch \\ Size}} & \multicolumn{1}{l}{\makecell[l]{Params \\ (Mb)}} & \multicolumn{1}{l}{\makecell[l]{Ratio- \\ to-2D}} & \multicolumn{1}{l}{\makecell[l]{FLOPs \\ (Gbps)}} & \multicolumn{1}{l}{\makecell[l]{Ratio- \\ to-2D}} &
    \multicolumn{1}{l}{\makecell[l]{Infer. \\ Speed \\ (imgs/s)}}\\
    \midrule
    2D  & 64\textcolor{red}{$\times$}64 & 3.3  & 1.0\textcolor{red}{$\times$} & 0.179 & 1.0\textcolor{red}{$\times$} & \underline{\textbf{419.2}}\\
    2D+TL & 64\textcolor{red}{$\times$}64 & 15.8  & 4.8\textcolor{red}{$\times$} & 2.51 & 14.0\textcolor{red}{$\times$} & 322.4\\
    \hdashline
    2.5D    & 64\textcolor{red}{$\times$}64\textcolor{red}{$\times$}9 & 29.8 & 9.0\textcolor{red}{$\times$} & 1.61 & 9.0\textcolor{red}{$\times$} & 103.2\\
    2.5D+TL & 64\textcolor{red}{$\times$}64\textcolor{red}{$\times$}9 & 141.9 & 43.0\textcolor{red}{$\times$} & 22.6 & 126.3\textcolor{red}{$\times$} & 55.5\\
    3D  & 64\textcolor{red}{$\times$}64\textcolor{red}{$\times$}64 & 25.7 & 7.8\textcolor{red}{$\times$} & 17.8 & 99.4\textcolor{red}{$\times$} & 207.9\\
    \hdashline
    2.75D   & 32\textcolor{red}{$\times$}123 & \underline{\textbf{3.1}}  & \underline{\textbf{0.9\textcolor{red}{$\times$}}} & \underline{\textbf{0.17}} & \underline{\textbf{0.95\textcolor{red}{$\times$}}} & 378.7\\
    2.75D+TL  & 32\textcolor{red}{$\times$}123 & 15.5 & 4.7\textcolor{red}{$\times$} & 2.35 & 13.1\textcolor{red}{$\times$} & 341.8\\
    \hdashline
    \multicolumn{1}{l}{2.75D\textcolor{red}{$\times$}3} & 32\textcolor{red}{$\times$}123\textcolor{red}{$\times$}3 & 9.3 & 2.8\textcolor{red}{$\times$} & 0.51 & 2.85\textcolor{red}{$\times$} & 224.7\\
    \multicolumn{1}{l}{2.75D\textcolor{red}{$\times$}3+TL}& 32\textcolor{red}{$\times$}123\textcolor{red}{$\times$}3 & 46.5 & 14.1\textcolor{red}{$\times$} & 7.06 & 39.4\textcolor{red}{$\times$} & 168.9\\
    \bottomrule
  \end{tabular}
  \label{table:cpm_effic}
\end{table}

\subsubsection{Backbone CNN}\label{strategy:backbone}
For a fair comparison, we kept the backbone CNN structure identical, only toggling between the 2D functions and their corresponding 3D functions (\eg convolution, pooling, Fig.~\ref{subfig:3D}). The backbone CNN structure we used is similar to proposed in~\citep{setio2016pulmonary} but deeper, to adequately extract discriminative features in different strategies, especially 3D. 

It consists of 3 consecutive double convolutional layers and max-pooling layers (see Fig.~\ref{subfig:2D}). The first double convolutional layers consist of two sets of 24 kernels (size is 3\textcolor{red}{$\times$}3\textcolor{red}{$\times$}1 and 3\textcolor{red}{$\times$}3\textcolor{red}{$\times$}24 each). Then Max-pooling is used in the pooling layer, which down-samples the patch size by half (\eg from 24@64\textcolor{red}{$\times$}64 to 24@32\textcolor{red}{$\times$}32 after the first max-pooling layer) by taking maximum values in non-overlapping windows of size 2\textcolor{red}{$\times$}2 (stride of 2). The second two convolutional layers consist of 48 kernels of size 3\textcolor{red}{$\times$}3\textcolor{red}{$\times$}24 and 3\textcolor{red}{$\times$}3\textcolor{red}{$\times$}48, respectively. The third double convolutional layer consists of two sets of 96 kernels with sizes of 3\textcolor{red}{$\times$}3\textcolor{red}{$\times$}48 and 3\textcolor{red}{$\times$}3\textcolor{red}{$\times$}96. Each kernel outputs a 2D feature map (\eg 24 of 64\textcolor{red}{$\times$}64 images after the first convolutional layer, which is denoted as 24@64\textcolor{red}{$\times$}64 in Fig.~\ref{subfig:2D}). The last layer is a fully connected layer with 512 output units. The Rectified linear units (ReLU) activation function is applied in each convolutional layer and fully connected layer, where the activation for a given input is obtained as $a = max(0, x)$.


\subsubsection{2D}\label{strategy:2D}
As shown in Fig.~\ref{subfig:2D}, a single 2D slice of size 64\textcolor{red}{$\times$}64 pixels is extracted on the X-Y plane at the center of the Z axis from each VOI. A single 2D CNN stream is applied. the backbone network architecture was described in Section \ref{strategy:backbone}, followed by a fully connected layer prior to the final binary classification layer.

\subsubsection{2.5D}\label{strategy:2.5D}
In this experiment, the multi-view CNN architecture we used is similar to the proposed in~\citep{setio2016pulmonary} with a deeper backbone (see Fig.~\ref{subfig:2.5D}). We adopted their optimized hyperparameters as well (i.e., learning rate, number of views, fusion method). The input of the network is a 64\textcolor{red}{$\times$}64\textcolor{red}{$\times$}9 patch, which consists of nine 2D views of the 3D volume as shown in Fig.~\ref{subfig:2.5D}. All nine 2D views are fed into 2D CNNs in parallel streams, the outputs of which are then fused for a final binary decision. Additionally, multiple 2D CNNs need to be fused to generate the final classification result. In this experiment, late fusion \citep{setio2016pulmonary} was implemented for comparison with other strategies. The late-fusion method \citep{prasoon2013deep,karpathy2014large} concatenates each of the 512 output units from all 2D CNNs and fully connects the concatenated outputs directly to the classification layer (see Fig.~\ref{subfig:2.5D}). By combining the information from multiple views, this strategy has the potential to learn 3D characteristics. 

\subsubsection{3D}\label{strategy:3D}
Besides the 2D and 2.5D strategies, it is a frequently adopted strategy to directly use the full 3D volume as input to CNNs for nodule classification (\eg 3D-CNN \citep{alakwaa2017lung}). In this experiment, the input is a 64\textcolor{red}{$\times$}64\textcolor{red}{$\times$}64 pixel patch. The used 3D CNN architecture for comparison (See Fig.\ref{subfig:3D}) is the same layer structure of the 2D CNN described in Section \ref{strategy:2D}, except that \textcolor{red}{2D convolution layer (Conv2D) and 2D Max-Pooling layer are replaced with 3D convolution layer (Conv3D) and 3D Max-Pooling layer respectively and filters are changed from 2D to 3D correspondingly.}

\subsubsection{Our proposed 2.75D and 2.75D\textcolor{red}{$\times$}3}\label{strategy:2.75D}
The proposed 2.75D strategy differs from 2.5D and 3D strategies essentially in the way of representing 3D information. As shown in Fig.~\ref{subfig:2.75D}, a 2D image of size 32\textcolor{red}{$\times$}123 pixel was extracted from each 3D VOI of size 64\textcolor{red}{$\times$}64\textcolor{red}{$\times$}64 pixel by applying the spiral scanning technique as described in Section \ref{sec:method}. Such 2D patches were fed into the 2D CNN model, which shares the same layer architecture as described in Section \ref{strategy:2D}.

We also explore a multi-view 2.75D strategy to boost the advantage of 2.75D. A three-view image of size 32\textcolor{red}{$\times$}123\textcolor{red}{$\times$}3 \textcolor{red}{pixels} was extracted from each 3D VOI by applying the spiral scanning technique from three different axes. Similar to the 2.5D CNN model, multi-view 2.75D also used multiple 2D CNNs with the late fusion (see Fig.~\ref{subfig:2.75Dx3})

\begin{table*}[htb]
\setlength\tabcolsep{1.5pt}
\scriptsize
  \centering
  \caption{Result comparison among all five strategies based on 10-fold (lung CT and Breast MRI) and 5-fold (Prostate MRI) cross-validation. We also calculated the significance of the difference between our proposed strategies and other \textcolor{red}{methods}. Superscript $^{1}$ means there is a significant improvement (p$<$0.05) between our methods and 2D (or between our methods in transfer learning and 2D TL), $^{2}$ shows comparing with 2.5D or 2.5D TL our methods have a significant advantage (\eg{ 2.75D/\textcolor{red}{2.75D$\times$3} v.s. 2.5D or 2.75D TL/2.75D\textcolor{red}{$\times$}3 TL v.s.2.5D TL}) and $^{3}$ indicate that the results of our method are significantly higher than 3D.}
  \begin{tabular}{lll:ll:ll}
    \toprule
    & \multicolumn{2}{c}{Lung CT} &\multicolumn{2}{c}{Breast MRI} & \multicolumn{2}{c}{Prostate MRI} \\
    \cmidrule(r){2-3}  \cmidrule(rl){4-5} \cmidrule(rl){6-7}
     Strategy & \multicolumn{1}{c}{CPM} & \multicolumn{1}{c}{AUC} & \multicolumn{1}{c}{CPM} & \multicolumn{1}{c}{AUC} & \multicolumn{1}{c}{CPM} & \multicolumn{1}{c}{AUC} \\
    \midrule
    2D & 0.473 [0.444-0.504] & 0.910 [0.899-0.921] & 0.815 [0.778-0.852] & 0.967 [0.956-0.977] & 0.357 [0.244-0.472] & 0.540 [0.452-0.625]\\
    2.5D & 0.636 [0.605-0.669] & 0.940 [0.932-0.949] & 0.875 [0.844-0.903] & 0.971 [0.960-0.981]& 0.453 [0.317-0.596] &  0.635 [0.563-0.710]\\
    3D & 0.714 [0.684-0.745] & 0.958 [0.950-0.966] & 0.869 [0.835-0.900] & 0.971 [0.960-0.981] & 0.467 [0.348-0.590] &  0.660 [0.587-0.736]\\
    2.75D & 0.692 [0.660-0.723]$^{1,2}$ & 0.955 [0.947-0.964]$^{1,2}$ & 0.868 [0.838-0.898]$^{1}$ & \underline{\textbf{0.975 [0.965-0.983]$^{1}$}} 
    & 0.502 [0.378-0.622]$^{1}$ &  0.663 [0.590-0.733]$^{1}$\\
    2.75D\textcolor{red}{$\times$}3 & \underline{\textbf{0.732 [0.702-0.761]$^{1,2}$}} & \underline{\textbf{0.964 [0.956-0.971]$^{1,2}$}} & \underline{\textbf{0.879 [0.849-0.907]$^{1}$}} & 
    0.974 [0.964-0.982]$^{1}$ & \underline{\textbf{0.545 [0.420-0.668]$^{1}$}} &  \underline{\textbf{0.669 [0.588-0.746]$^{1}$}}\\
    \hdashline
    2D+TL & 0.598 [0.568-0.629] & 0.940 [0.929-0.950] & 0.839 [0.804-0.872] & 0.964 [0.951-0.975] & 0.378 [0.271-0.493] &  0.562 [0.488-0.633]\\
    2.5D+TL & 0.740 [0.712-0.769] & 0.963 [0.956-0.970] & 0.883 [0.852-0.912] & 0.972 [0.961-0.981] &  0.495 [0.363-0.624] & 0.644 [0.564-0.724]\\
    2.75D+TL & 0.767 [0.736-0.796]$^{1,2,3}$ & 0.975 [0.969-0.981]$^{1,3}$ & 0.889 [0.860-0.915]$^{1}$ & 0.980 [0.971-0.987]$^{1,2,3}$ & \underline{\textbf{0.531 [0.410-0.657]$^{1}$}} &  0.681 [0.608-0.754]$^{1}$\\
    2.75D\textcolor{red}{$\times$}3+TL & \underline{\textbf{0.795 [0.768-0.822]$^{1,2,3}$}} & \underline{\textbf{0.982 [0.977-0.987]$^{1,2,3}$}} & \underline{\textbf{0.896 [0.867-0.923]$^{1,3}$}} & \underline{\textbf{0.981 [0.972-0.988]$^{1,2,3}$}} &  0.522 [0.407-0.641]$^{1}$ & \underline{\textbf{0.692 [0.621-0.759]$^{1}$}}\\
    \bottomrule
  \end{tabular}
  \label{table:cpm_auc_strategies}
\end{table*}

\begin{figure*}[htb]
  \centering
  \includegraphics[clip, trim=0.4cm 0.3cm 0.4cm 0.3cm, width=.99\linewidth]{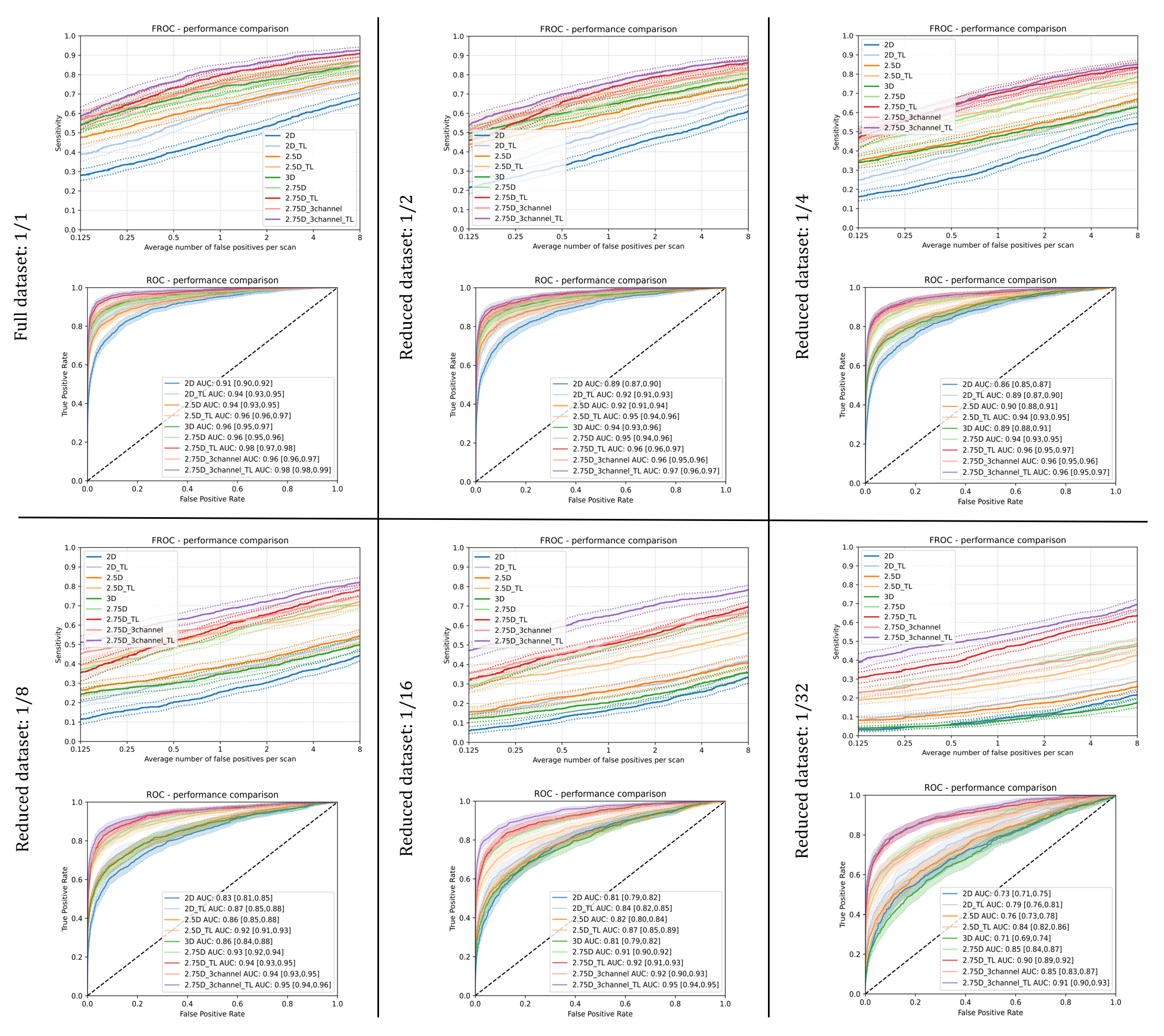}
\caption{FROC and ROC curves of various strategies on reduced datasets.}
\label{fig:luna}
\end{figure*}

\subsection{Impact of Transfer Learning on 2.75D and 2.75D\textcolor{red}{$\times$}3}
A major advantage of 2.75D is that it allows broader access to the power of transfer learning (TL). There are numerous existing pre-trained models based on large 2D datasets, \eg ImageNet, whereas the amount of existing pre-trained 3D models are scarce. Owing to the fact that medical data acquisition requires privacy consent and expensive hospital resources, data are often the bottleneck for model robustness and generalizability. To address the limitations of clinical data, transfer learning has been increasingly exploited in the medical field \citep{pan2010survey,wang2017hospital,rajpurkar2017chexnet} to attain state-of-the-art performance.

In this work, we assessed the performance impact of transfer learning on the proposed 2.75D strategy. 3D volumes were first transformed into 2D images (by 2D, 2.5D, 2.75D, or 2.75D\textcolor{red}{$\times$}3 methods, respectively), which were fed into the \textcolor{red}{pre-trained} VGG network on ImageNet for fine-tuning. The first three blocks were frozen during fine-tuning. As for 2.5D and 2.75D\textcolor{red}{$\times$}3, we use the same late fusion method as the above CNN architecture. For fairness, We not only compare the performance of our proposed methods in the fine-tuned model with the aforementioned three strategies (2D, 2.5D, and 3D) but also compare with 2D and 2.5D with transfer learning. 

\subsection{Evaluation of Efficiency}
Another clear advantage of our strategies is efficiency. To evaluate the efficiency, we computed the model size and computational cost expressed in terms of Parameters (Mb), and floating point operations (FLOPs (Gbps)), respectively (see Table \ref{table:cpm_effic}). We also counted the inference time to calculate the inference speed (images/s). Note that, in order to perform a fair comparison, the inference time includes resampling a 64\textcolor{red}{$\times$}64\textcolor{red}{$\times$}64 cube to a 2D, 2.5D, 2.75D, or 2.75D\textcolor{red}{$\times$}3 image.

\subsection{Impact of Limited Data}
In comparison to 3D CNN, a comparable 2D CNN architecture possesses several times smaller trainable parameters. As 2.75D compressed 3D information into 2D, we hypothesize that the advantage of 2.75D would be more prominent in case of limited training data. Note that, due to the limited amount of images in the Breast MRI and Prostate MRI dataset, this experiment was only conducted using the LIDC-IDRI dataset. In this experiment, we evaluated the performance of four strategies with various training \textcolor{red}{data sizes}, while keeping the test set intact. The training set was downscaled by a factor of 2, 4, 8, 16, and 32, while keeping the \textcolor{red}{positive-to-negative sample} ratio unchanged. Using such downscaled training sets, four aforementioned CNN strategies and three transfer learning strategies were evaluated.

\subsection{Implementation Details}
\textcolor{red}{One of the challenges of using CNNs is to efficiently optimize the model parameters given the training dataset. To limit the differences between all the CNNs in this study, the same batch size of 64, loss function measured by cross-entropy error, $1\times10^{-5}$ of the initial learning rates, and Adam optimizer \citep{kingma2014adam} with 0 weight decay were shared among all learning strategies. We monitored the training through the metric of accuracy on the validated set and \textcolor{red}{saved} the best model. Early stopping with maximum of 50 epochs was also applied to all CNN architectures when the validation accuracy does not increase for \textcolor{red}{ten} consecutive epochs. We adopted normalized initialization for the model weights as proposed by Glorot \etal~\cite{glorot2010understanding}. The biases were set initially to zero. For transfer learning, we also applied the same hyperparameter and training strategies. Ten/five-fold cross-validation was used to train and evaluate models. In the ten-fold cross-validation, for each fold, we used seven subsets for training, two subsets for validation, and one subset for testing. In the five-fold cross-validation, for each fold, we used three subsets for training, one subset for validation, and one subset for testing. All the experiments were performed using a single NVIDIA RTX A6000 with 48GB VRAM.}

\subsection{Evaluation Metrics and Statistical Analysis}
The performances of the four strategies were compared based on two metrics: Area Under the Receiver Operating Characteristic Curve (AUC) and Competition Performance Metric (CPM) \citep{niemeijer2010combining}. AUC is a commonly used evaluation metric for machine learning tasks, which represents the classification performance by the area under the Receiver Operating Characteristic curve (ROC). We used the paired DeLong’s test \citep{sun2014fast} to assess the significance between the proposed strategy and others. CPM is defined as the average sensitivity at seven operating points of the Free-response ROC curve (FROC): 1/8, 1/4, 1/2, 1, 2, 4, and 8 false positives per scan. We also calculated the significance of the difference between two FROC curves, as detailed in \citep{dalmics2018fully}.
In the lesion diagnostic classification task for the prostate dataset, the FROC is defined as the average sensitivity at four operating points: 1/8, 1/4, 1/2, and 1 false positive per scan. A larger false positive rate is meaningless due to a different proportion of positive and negative samples.

\section{Results}\label{sec:experimental_results}

\subsection{Comparison of Efficiency}
As shown in Table~\ref{table:cpm_effic}, Notably, the patch size (number of pixels) of 2.75D is even smaller than that in 2D. Therefore, the parameters and FLOPs of the 2.75D is the smallest. The 3-channel 2.75D is also smaller than the 2.5D and 3D approach by multiple factors of magnitude in model size and computational cost. In terms of inference speed, the proposed strategy is still faster than 2.5D and 3D, even including the time of spirally representing 3D patches to 2D images. Moreover, this advantage in efficiency still holds true in the case of transfer learning. 

\subsection{Comparison of Classification Performance}
Table~\ref{table:cpm_auc_strategies} shows the CPM and AUC scores of classification results on three different datasets using four different strategies. Fig.~\ref{fig:luna} shows the FROC curve of pulmonary nodule classification on the LIDC-IDRI dataset. The CPM and AUC scores of 2.75D\textcolor{red}{$\times$}3 are even higher than the scores of the 3D (AUC: 0.958, CPM: 0.714). In breast cancer classification and prostate lesion classification tasks, 2.75D strategies demonstrate significant outperform 2D. Notably, 2D and 2.75D share almost the same patch size in the number of pixels, whereas 2.75D achieves a CPM of 0.73, which is 55\% higher than that of 2D. The proposed 2.75D strategy demonstrates its capability and efficiency in capturing spatial 3D information.


\subsection{Impact of Transfer Learning}
As shown in Table~\ref{table:cpm_auc_strategies} and Fig.~\ref{fig:luna}, the best performance is achieved using the 2.75D strategy on the top of fine-tuned VGG16 models, with a significant gain in both CPM and AUC versus the 3D strategy on both lung CT and Breast MRI datasets. More importantly, our methods maintained or even expanded their significant advantages compared with 2D and 2.5D strategies in the case of transfer learning. This experiment shows the potential of 2.75D for further performance boost with its compatibility to transfer learning techniques. In contrast, currently, it is challenging to find any publicly available pre-trained 3D model trained from a dataset with a comparable data-size to ImageNet. 


\subsection{Evaluation under Limited Data}
Fig.~\ref{fig:luna} shows the comparison between different strategies when trained on the reduced dataset by a factor of 2, 4, 8, 16, and 32, respectively. As data size reduces, the advantages of 2.75D and fine-tuned 2.75D strategy become more prominent. That is to say, 2.75D and 2.75D\textcolor{red}{$\times$}3 represent the 3D volume in an effective way, which enables the utilization of 2D CNNs with relatively less trainable parameters than 3D CNNs. This is especially beneficial when the training data size is limited, which is a typical situation in the medical imaging field. Interestingly, when training sample have been reduced by a factor of 16, both 2.5D and 2D approaches started to outperform the 3D approach, which again shows that a proper 3D approach requires sufficient data. With less training data, the 2.75D strategy can be the first choice.


\section{Discussion and Conclusion}
In this study, we have presented 2.75D, as a strategy to efficiently and effectively represent 3D features in 2D for 2D CNN analytical tasks, especially in medical imaging. 
\textcolor{red}{In terms of performance, our approaches substantially outperform the traditional 2D method in different medical imaging tasks on three datasets. Although the 2.75D methods may not always significantly outperform 2.5D or 3D in all circumstances, 2.75D has shown its advantage when training data is relatively small.}
In terms of computational efficiency, the 2.75D approach is equivalent to the 2D approach and even the 2.75D$\times$3 approach is one-third of 2.5D. The inference times are also several times shorter than the 2.5D and 3D approaches. 
Furthermore, the 2.75D strategy not only eases the adoption of the vast amount of pre-trained 2D deep learning models for transfer learning for medical imaging but also significantly reduces required model parameters. 
Remarkably boosted performance has been shown with transfer learning, as well as in the case of limited training data.

This is the opening research on representing 3D volumetric data in 2D to boost classification efficiency while preserving 3D information.
To explore a better 2D representation of 3D volumes, we revisited the spiral scanning technique in this work. The spiral sampling can be utilized to map voxels in the 3D space onto a 2D plane while retaining the contextual correlation between the target voxels, particularly the spatial correlation between texture features. Thus, the basic 2D classification method with spiral sampled images as input can perform well in 3D nodal classification in a more efficient way.

Compared with the 2.5D method, our 2.75D methods could globally incorporate richer volumetric information of neighboring voxels with geometry relation preserved, which is more efficient than the 2.5D method. 
However, the 2.5D method locally represents 3D volumes by extracting multiple 2D patches that are only sparsely sampled in several limited predefined orientations.
The voxels from 3D are not strategically sampled in a single 2D image and geometry relations between voxels in 3D are lost. 
Therefore, when reducing the data size, the advantages of 2.75D and fine-tuned 2.75D strategy become more prominent. 
As for the 3D method, given sufficiently large data, a tailored 3D CNN may theoretically outperform 2.75D. However, a rich annotated 3D dataset is often not easy to obtain in the medical domain. This is where 2.75D can contribute.


For future work, there are multiple directions. From our experiments, we showed the effectiveness of spiral sampling for the lung cancer detection task. The benign tumours, in general, are more ball-like or oval, while malignant tumours can have very irregular shapes. Our 2.75D could more densely retain information about the inside and boundary of the tumours than the 2.5D and 2D methods. One of the remaining questions is how robust our methods are with respect to different sizes or shapes of 3D objects \eg organs, tumours, tissues, cells, and vessels in medical imaging. We will also investigate how the different arrangements of the sampling trajectory affect the performance in future work. As part of future work, we will further verify the effectiveness of the proposed 2.75D strategy in various deep learning tasks in medical imaging.


In conclusion, we have presented a novel CNN strategy by representing 3D images into 2D features. With extensive experiments on three public data sets, the potential of 2.75D in achieving superior and faster performance has been revealed. 
 We believe that our work adds new perspectives to the CV community. 
From a green AI perspective, our generalizable approach shows great potential in 3D medical image analysis, considering the scarcity of annotated medical data.

\section{Acknowledgment}
Xin Wang is funded by Chinese Scholarship Council scholarship (CSC).


\bibliography{cas-refs}

\end{document}